\documentclass[aps,amsmath]{revtex4}
\usepackage{graphicx}

\renewcommand{\d}{\partial}
\newcommand{\nn}{\nonumber\\}

\newcommand{\rh}{\varrho}
\newcommand{\exv}[1]{\left\langle{#1}\right\rangle}
\newcommand{\exvs}[1]{\langle{#1}\rangle}
\newcommand{\ep}{\varepsilon}

\newcommand{\q}{{\bf q}}
\renewcommand{\k}{{\bf k}}
\newcommand{\p}{{\bf p}}

\newcommand{\Tr}{\mathop{\textrm{Tr}}}

\newcommand{\pint}[2]{{\int\!\frac{d^{#1}#2}{(2\pi)^#1}\,}}
\newcommand{\pintz}[1]{{\int\!\frac{d #1}{2\pi}\,}}

\newcommand{\bra}[1]{\left\langle{#1}\right|}
\newcommand{\ket}[1]{\left|{#1}\right\rangle}

\newcommand\lsim{\mathrel{\rlap{\lower4pt\hbox{\hskip1pt$\sim$}} \raise1pt\hbox{$<$}}}                
\newcommand\gsim{\mathrel{\rlap{\lower4pt\hbox{\hskip1pt$\sim$}} \raise1pt\hbox{$>$}}}                

\begin{document}

\title{Non-universal lower bound for the shear viscosity to entropy
  density ratio}
\author{Antal Jakovac}
\affiliation{Physics Institute, BME Technical University, H-1111 Budapest, Hungary}
\date{\today}

\begin{abstract}
  The lower bound of the shear viscosity to entropy density ratio is
  examined using an exact representation of the ratio through the
  density of states. It is shown that the lower bound in a generic
  physical system is not universal, its value is determined by the
  entropy density. Some examples of physical systems are discussed in
  the paper where one can expect violation of the conformal $1/4\pi$
  value.
\end{abstract}

\maketitle

\section{Introduction}

Experimental evidence from RHIC \cite{Adler:2003kt,Adams:2003am}
suggests \cite{Shuryak:2003xe,Teaney:2003kp} that the QCD matter near
the (would-be) critical temperature is very close to an ideal fluid,
the shear viscosity to entropy density ratio is $\eta/s < 0.2$
\cite{Romatschke:2007mq,Heinz:2008qm}. Such small values suggest very
strong interactions, and it is therefore a big theoretical challenge
to explain the actual $\eta/s$ in this system. Direct calculations
with weak coupling resummed perturbation theory in various models
\cite{Jeon:1994if,Wang:2002nba, Aarts:2005vc,Huot:2006ys,
  Gagnon:2007qt} and resummations with Boltzmann-equations
\cite{Jeon:1995zm, Arnold:2002zm} suffer generally from the problem
that the unperturbed system, ie. the free gas has infinite viscosity,
and so small perturbations necessarily remain in the large viscosity
regime. This is reflected in the result $\eta/s\sim 1/(g^4\log g)$
which is large if the coupling is small. At stronger coupling
subleading corrections may alter this result: for example the
inclusion of $2-3$ Bremsstrahlung processes into the Boltzmann
equation results in a significant difference near the critical region
of QCD \cite{Xu:2007ns,Xu:2007jv}.

Since perturbation theory is not well controlled in the strong
coupling regime, one has to look for alternative methods. A
potentially exact method is lattice Monte Carlo simulations. However
the actual measurement of the transport coefficients is plagued by the
necessity of analytic continuation from imaginary to real time based
on a discrete set of data. The inversion of the integral equation for
the analytic continuation then has very little sensitivity to the
important infrared physics \cite{Petreczky:2007js}. Therefore existing
MC results \cite{Meyer:2007ic} depend on not fully controlled
assumptions.

A completely different strategy is when one uses a dual description of
the theory, where the strong coupling limit is mapped to the small
coupling limit of the dual theory. In case of QCD-like theories this
is possible only with a much larger symmetry group, the ${\cal N}=4$
SYM theories, where supersymmetry together with the holographic
principle makes possible to work, at large $N_c$ and large t'Hooft
coupling, in a five dimensional gravity with AdS metric
\cite{Aharony:1999ti}. There the shear viscosity of the conformal
field theory can be calculated from graviton absorption
\cite{Kovtun:2004de}, resulting in the celebrated $1/4\pi$. In a large
class of AdS models this seems to be a lower bound \cite{Myers:2008yi,
  Myers:2008me}.

Although ${\cal N}=4$ SYM is not QCD, one can argue independently that
there is a lower bound in QCD itself \cite{Danielewicz:1984ww}. Then
the value of this lower bound may come universally from the large
$N_c$ large t'Hooft coupling supersymmetric case \cite{Kovtun:2004de}.

The statement that $1/4\pi$ would be the lower bound in any system for
the $\eta/s$ ratio is, however, not really proven, only argued for,
and it is supported by the fact that we never encountered systems
where this bound was violated. But the argumentation has weak points,
which suggest that it is not so general as it is commonly
believed. First, in the gravity side one can find higher curvature AdS
models, where this conclusion fails \cite{Kats:2007mq, Buchel:2008vz,
  Cai:2009zv}. In the framework of nonrelativistic gases one can also
construct counterexamples \cite{Cohen:2007qr,Cherman:2007fj}. Although
these examples may describe metastable states \cite{Son:2007xw}, but
the question of the applicability range of the result is not answered
reassuringly.

More importantly, the argumentation is strongly based on the
quasiparticle picture. The generic argumentation says
\cite{Kovtun:2004de} that $\eta/s \sim E \tau$ where $E$ is the energy
of an excitation, $\tau$ is the lifetime, and uncertainty principle
states that $E\tau\gsim\hbar$ which yields $\eta/s\gsim\hbar$. But
uncertainty principle, in fact, states that the \emph{uncertainty} of
the energy measurement is related to the lifetime, not the energy
value itself. In quasiparticle picture the uncertainty cannot be
larger than the energy value itself, but in a generic theory this may
not be the case. So, in fact the restriction $E\tau\gsim\hbar$ is not
a law of nature, but rather the applicability range of the
quasiparticle picture. If the small width quasiparticle approximation
cannot be applied, then we have no real arguments in favor of the
lower bound. Recently in \cite{NoronhaHostler:2008ju} the authors have
shown that the presence of a continuum in the particle spectrum has
important effect on the shear viscosity to entropy density ratio.

To clarify these issues one would need exact statements about the
lower limit of $\eta/s$ in physical systems. This is what is attempted
in this work, using first principles. For the derivation we have to
make some assumptions about some current matrix elements, then one can show
that if the transport is dominated by a finite number of quantum
channels (ie. there are no large number of particle species) then at
finite entropy density the $\eta/s$ ratio has a lower bound. This
lower bound, however, depends on the entropy itself; for small entropy
density case it is
\begin{equation}
  \frac{\eta}s\biggr|_{\mathrm{min}} \sim \frac1{N_Q}\,
  \frac{s}{T^3}\, \frac1{LT},
\end{equation}
where $N_Q$ is the number of dominant quantum channels, $L$ is the
``interaction range''. This latter can be defined so that parts of the
system separated more than $L$ do not interact, or interact weakly; in
other words in systems larger than $L$ the entropy is an extensive
quantity. For more details and for general $s$ formulae see Section
\ref{sec:min} and \eqref{etaperesmin}.

The mathematical reason why $\eta_{min}\sim s^2$ depends qualitatively
on the fact that, as it will be explained in the paper in detail, the
transport coefficients come from integrals depending quadratically on
the energy density of states $\rh$, while the entropy density, for low
values of $s$, depends linearly on $\rh$. Since these integrals have
positive kernels, we can treat them as some averaging. So, up to
normalization factors $\eta\sim\exv{\rh^2}$ and $s\sim\exv{\rh}$. This
implies that with this normalization $\eta\ge s^2$, and the minimum is
therefore proportional to $s^2$.

In a finite degree of freedom quasiparticle system the entropy density
is proportional to $T^3$ and the interactions are screened at most at
a scale of the temperature, and so the right hand side is
constant. The value of this constant cannot be determined from a
general argumentation, but by the argumentation of
\cite{Danielewicz:1984ww,Kovtun:2004de} it may be $1/4\pi$. But as a
mathematical statement we can claim that there is no lower bound for
the shear viscosity to entropy density ratio.

The above formula also tells us where can we expect to have small
$\eta/s$ ratio: if the number of dominant quantum channels is large
(cf. \cite{Cohen:2007qr,Cherman:2007fj}), if the system interacts on
very large scales (ie. large width systems, where the width is not
connected to the temperature), or if we are at low temperatures and
the entropy density vanishes faster than $T^3$. 

In the paper we discuss some systems, where possibly violations of the
$1/4\pi$ value are present. These are systems with strong wave
function renormalization, large width finite temperature systems and
low temperature non-quasiparticle systems. All of these cases can
appear in real systems: strong wave function renormalization shows up
in strongly coupled quasiparticle systems; at finite temperature QCD
we expect large width systems \cite{Peshier:2005pp}. Finally in
systems where there are zero mass excitations, the spectral function
must not have a separated Dirac-delta peak even at zero
temperature.

The structure of the paper is the following. We first derive exact
expressions for the Kubo formula and for the entropy (Section
\ref{sec:curcur}) through the density of states. In Section
\ref{sec:min} we study the mathematical structure of the $\eta/s$
ratio and determine its minimum. Next we study the quasiparticle-like
systems and discuss the small-width quasiparticle case in Section
\ref{sec:QP}. In Section \ref{sec:offshell} we examine different
situations where the excitations are not small-width quasiparticles,
and discuss the impact of the off-shell effects on transport. We
finish with conclusions in Section \ref{sec:concl}.

\section{The current-current correlator and the entropy density}
\label{sec:curcur}

In this section we derive exact representations for the Kubo formula
and the entropy density using the density of states (or energy
spectral functions).

\subsection{Kubo formula}

The shear viscosity, the linear response coefficient for the
transversal momentum difference in a plasma, is defined through the
Kubo formula in the framework of quantum field theory. We introduce
\begin{equation}
  C(x) = \exv{[\pi_{xz}(x),\pi_{xz}(0)]},\qquad \pi_{ij} = T_{ij} -
  \frac13 \delta_{ij} T_\ell^\ell,\qquad i,j=1,2,3
\end{equation}
where $T$ is the energy-momentum tensor of the system, and the
expectation value has to be taken in the equilibrium system:
\begin{equation}
  \exvs{\hat Q} = \frac1{\cal Z} \Tr e^{-\beta H}\hat Q,\qquad {\cal
    Z} = \Tr e^{-\beta H}.
\end{equation}
The viscosity is obtained after Fourier transformation as
\begin{equation}
  \eta = \lim\limits_{p_0\to 0} \frac{C(p_0,\p=0)}{p_0}.
\end{equation}
This discussion can be generalized by using instead of $\pi_{ij}$ a
general conserved current $J_i$, and look at the linear response
function
\begin{equation}
  C_J(x) = \exv{[J_i(x),J_i(0)]},
\end{equation}
(there is no summation on indices $i$). Because of rotational
invariance of the equilibrium system, only the diagonal elements are
nonzero here. The corresponding transport coefficient will be called
$\eta_J$.

We continue by inserting energy eigenstates into the expectation value
above.
\begin{eqnarray}
  C_J(x) && = \frac1{\cal Z} \sum_{n,m} \biggl\{\exv{n|e^{-\beta H} J_i(x)|m}
    \exv{m|J_i(0)|n} - \exv{m|e^{-\beta H} J_i(0)|n}
    \exv{n|J_i(x)|m}\biggr\} =\nn&&= \frac1{\cal Z} \sum_{n,m} \left(
      e^{-\beta E_n} - e^{-\beta E_m}
    \right)\,\exv{n|J_i(x)|m}\exv{m|J_i(0)|n}.
\end{eqnarray}
The possible states in the system is characterized by their total
energy $k_0$ total momentum $\k$ and other quantum numbers ${\cal K}$
\begin{equation}
  \ket n = \ket{k_0,\k,{\cal K}}.
\end{equation}
Similarly we denote $\ket m=\ket{q_0,\q,{\cal Q}}$. We will also use
the four-momentum notation $k=(k_0,\k)$ and $q=(q_0,q)$.

In a realistic quantum field theory the allowed energy eigenvalues
can form a continuum, and at finite temperature this is always the
case. Therefore it is more adequate to write the sum over the energy
in the above expression as an integral over a spectral density. In
finite volume spatial momenta form a discrete set, but at larger
volumes a continuum representation is applicable for them, while other
quantum numbers remain discrete:
\begin{equation}
  {\bf 1}=\sum_n \ket{n}\bra{n} = \sum_{\k,{\cal K}}\pintz{k_0} \, \rh_{\cal K}(k)
  \;\ket{k,{\cal K}} \bra{k,{\cal K}} = V \sum_{\cal K}\pint4k \, \rh_{\cal K}(k)
  \;\ket{k,{\cal K}} \bra{k,{\cal K}}.
\end{equation}
Since the state $\ket{k,{\cal K}}$ is dimensionless in our
normalization, therefore $\rh$ has the dimension of inverse
energy. With the energy density of states we can write
\begin{equation}
  C_J(x) = \frac{V^2}{\cal Z} \sum_{\cal K} \pint4k \rh_{\cal K}(k)
  \sum_{\cal Q} \pint4q \rh_{\cal Q}(q) \,\left( e^{-\beta k_0} -
    e^{-\beta q_0}\right) \exv{k,{\cal K}|J_i(x)|q,{\cal Q}}
  \exv{q,{\cal Q}|J_i(0)|k,{\cal K}}.
\end{equation}

Using the generator of the space-time translation we write
\begin{equation}
\label{Jmatel}
  \exv{k,{\cal K}|J_i(x)|q,{\cal Q}} = e^{i(k-q) x} \exv{k,{\cal
      K}|J_i(0)|q,{\cal Q}},
\end{equation}
then the Fourier transformation yields
\begin{equation}
  C_J(p) = \frac{V^2}{\cal Z} \sum_{\cal K} \pint4k \rh_{\cal K}(k)
  \sum_{\cal Q} \pint4q \rh_{\cal Q}(q) \,(2\pi)^4\delta^{(4)}(k-q+p) \,
  \left( e^{-\beta k_0} - e^{-\beta q_0}\right) |\exv{k,{\cal
      K}|J_i(0)|q,{\cal Q}}|^2.
\end{equation}
Then the transport coefficient reads
\begin{equation}
  \eta_J = \beta \, \frac{V^2}{\cal Z} \sum_{{\cal K},{\cal Q}} \pint4k
  \rh_{\cal K}(k)\,\rh_{\cal Q}(k) e^{-\beta k_0} |\exv{k,{\cal
      K}|J_i|k,{\cal Q}}|^2.
\end{equation}
Since $J_i$ cannot change the quantum channel without making any
change in both energy and momentum, so nonzero result comes only from
${\cal K}={\cal Q}$:
\begin{equation}
  \eta_J = \beta \, \frac{V^2}{\cal Z} \sum_{\cal K} \pint4k \rh_{\cal
    K}^2(k)\, e^{-\beta k_0} |\exv{k,{\cal K}|J_i|k,{\cal K}}|^2.
\end{equation}

The (temperature-independent) matrix element of $J_i$ can be
calculated in the free theory with the result
\begin{equation}
  \exv{k|J_i|k}_0 = q\frac{k_i}{k_0V}
\end{equation}
where $q$ is the charge carried by the current. In case of the energy
momentum tensor $q=k_j$:
\begin{equation}
  \exv{k|T_{xz}|k}_0 = \frac{k_xk_z}{Vk_0}.
\end{equation}
In general, we expect an expression of similar structure, since the
external momentum carries the Lorentz index of the current, and by
dimensional reasons it must be divided by an energy-like
quantity. There can still be a Lorentz-invariant factor:
\begin{equation}
  \label{curform}
  \exv{k,{\cal K}|J_i|k,{\cal K}} = q\frac{k_i}{k_0V} {\cal J}_{\cal
    K}(k^2),\qquad\mathrm{or}\quad \exv{k,{\cal K}|T_{xz}|k,{\cal K}}
  = \frac{k_xk_z}{Vk_0} \,{\cal T}_{\cal K}(k^2).
\end{equation}
The reduced current matrix element ${\cal J}_{\cal K}$ depends on the
system under consideration, and we cannot tell much more on a generic
ground. We will assume, however, that ${\cal J}_{\cal K}$ does not
grow at large $k_0$ exponentially.

Since the only directional dependence is in the current, we can
average angular dependence using \eqref{curform}. Since
$\overline{\cos^2\theta} = 1/3$ and $\overline{\sin^2\theta
  \cos^2\theta \cos^2\phi} = 1/15$, we can write
\begin{equation}
\label{etaJ}
  \eta_J = \frac{\beta}{3\cal Z} \sum_{\cal K} \pint4k \,
  \frac{\k^2}{k_0^2} \, e^{-\beta k_0}\, ({\cal J}_{\cal K}(k)
  \rh_{\cal K}(k))^2,
\end{equation}
and for the shear viscosity
\begin{equation}
  \label{eta}
  \eta = \frac\beta{15 \cal Z} \sum_{\cal K} \pint4k \,
  \frac{(\k^2)^2}{k_0^2} \, e^{-\beta k_0}\, ({\cal T}_{\cal K}(k) \rh_{\cal K}(k))^2.
\end{equation}
The volume dependence canceled, and so we interpret these formulae as
infinite volume limit results.

\subsection{The entropy}

To calculate the entropy density we start by computing the free energy
as
\begin{equation}
  {\cal Z} = e^{-\beta F} = \Tr e^{-\beta H} = \sum_n e^{-\beta E_n} =
  \sum_{\k,{\cal K}} \pintz{k_0}\rh_{\cal K}(k) e^{-\beta k_0} = V
  \sum_{\cal K} \pint4k \rh_{\cal K}(k) e^{-\beta k_0}.
\end{equation}
In a strongly coupled system the volume dependence of the free energy
can be arbitrary. When we increase the linear size of the system,
after a certain scale $L$ one can observe more and more accurately the
the linear volume dependence, ie. the free energy becomes an extensive
quantity. That means that adjacent volume elements larger than $L$
interact (mostly) with surface interaction, while smaller volume
elements interact through the total volume. Therefore this
``interaction range'' serves also as an infrared cutoff for the
relevant interactions. In quasiparticle systems the interaction range
corresponds to the linear size of the cross section, the inverse of
free mean path. In general strongly interacting systems have large
$L$, weakly interacting ones have small $L$.

This is the scale, beyond which a coarse grained description is sensible
\cite{Langer:1967ax, Langer:1969bc}, since only there can one speak
about densities of different physical quantities. The free energy in a
coarse grained description is dominated by the integral of the free
energy density, while the corrections coming from the surface
interaction are subleading. Therefore in a microscopic description we
should choose the volume size at least $V=L^3$ in order to catch all the
relevant local physics effects. Therefore we define our free energy
density as
\begin{equation}
  f = -\frac T{L^3} \ln\left(L^3\sum_{\cal K} \pint4k \rh_{\cal
      K}(k)\,e^{-\beta k_0}\right).
\end{equation}
We can separate the ground state (vacuum) contribution, where always
\footnote{Note that in some systems there is a residual entropy at the
  vacuum. To avoid this case we should understand $s-s_{vac}$ for
  entropy density in the followings.}
$\rh_{vac}(k_0)=2\pi\delta(k_0)$:
\begin{equation}
  f = -\frac T{L^3} \ln\left(1+L^3\sum_{\cal K} \pint4k \rh_{\cal
      K}(k)\, e^{-\beta k_0}\right).
\end{equation}
From this generic form the entropy density reads
\begin{equation}
  s = \frac1{L^3} \ln\left(1+L^3 \sum\limits_{\cal K} \pint4k
    \rh_{\cal K}(k)\,e^{-\beta k_0}\right) +  \frac1{\cal Z}
  \sum\limits_{\cal K} \pint4k \,\beta k_0\, \rh_{\cal K}(k)\,e^{-\beta k_0}.
\end{equation}

Then the transport coefficient to entropy density ratio reads
\begin{equation}
\label{etaperes0}
  \frac{\eta_J}s = \frac{\displaystyle \frac\beta{3\cal Z}
    \sum_{\cal K} \pint4k \, \frac{\k^2}{k_0^2} \, e^{-\beta k_0}\,
    ({\cal J}_{\cal K}(k)\rh_{\cal K}(k))^2} {\displaystyle
    \frac1{L^3} \ln\left(1+L^3 \sum\limits_{\cal K} \pint4k
      \rh_{\cal K}(k)\,e^{-\beta k_0}\right) +  \frac1{\cal Z}
    \sum\limits_{\cal K} \pint4k \,\beta k_0\, \rh_{\cal K}(k)\,e^{-\beta k_0}}.
\end{equation}
In case of shear viscosity we have to substitute ${\cal J}^2 \to \k^2 {\cal
  T}^2/5$.

\section{Minimization of shear viscosity to entropy density ratio}
\label{sec:min}

Equation \eqref{etaperes0} gives the functional dependence of
$\eta_J/s$ on $\rh$, and so we can study mathematically the minimum of
this ratio by varying $\rh$. The above formula, however, contains yet
the unknown ${\cal J}$ factor. To say something constructive we have
to make assumptions on this function: we will assume that it can get
at most power law $k_0$ contributions in a strongly interacting
system. In this case we can perform an analysis to find the lower
bound.

There is a simple way to decrease the value of $\eta_J/s$: we start with
an arbitrary $\rh$ with finite $\eta_J/s$ and make a rescaling
$\rh\to\lambda\rh$ with $\lambda\to0$. Then the integrals of $\rh$
become small, we can make a linearization in $s$ with respect to
$\rh$. Then the numerator is quadratic, the denominator is linear, and
so $\eta_J/s\to\lambda \eta_J/s$, in the $\lambda\to0$ limit it
vanishes. 

In a physical system this simple procedure cannot be performed,
because we have a sum rule in each quantum channel
\begin{equation}
  \label{sumrule}
  \pintz{k_0} \rh_{\cal K}(k_0,\k) = U_{\cal K}(\k),\quad\forall\,
  {\cal K},\,\k.
\end{equation}
But this constraint is not really restricting. To understand it we
should remark that all $\rh$ integrals for $\eta_J$ and $s$ are
weighted by $e^{-\beta k_0}$. Therefore $\eta_J/s$ is sensitive to the
$k_0\le T$ regime, say, the infrared (IR) regime. The sum rule, on the
other hand, depends also on $k_0\ge T$ regime, the ultraviolet (UV)
regime. Therefore we may rescale the IR part of the energy density of
states and still maintain the sum rules.

To see this we assume that we can linearize the entropy density in
$\rh$ and try to find the minimal $\eta_J$. The generic case we will
discuss later. The assumption of linearizability also means that
${\cal Z}\approx 1$. To simplify the notation we introduce
\begin{equation}
  \sum\limits_{\cal K} \pint3\k = \int_Q.
\end{equation}
For the linearized case, in this symbolic notation the transport
coefficient and the linearized entropy density read
\begin{equation}
  \eta_J = \int_Q \pintz{k_0} \alpha_Q(k_0) \rh^2_Q(k_0) e^{-\beta
    k_0}, \qquad s =\int_Q \pintz{k_0} \gamma(k_0) \rh_Q(k_0)e^{-\beta
    k_0},
\end{equation}
where
\begin{equation}
  \alpha_Q(k_0) = \frac{\beta\k^2}{3k_0^2}\, {\cal J}^2_{\cal K}(k),
  \qquad \gamma(k_0) = 1+\beta k_0.
\end{equation}
We shall minimize $\eta_J$ with the constraint \eqref{sumrule}, and so
we apply Lagrange multipliers to each quantum channel. To have a
control to the thermodynamics we also fix the entropy density
$s$. Then we have to minimize the expression
\begin{equation}
  \min\int_Q\pintz{k_0} \bigl[ \alpha_Q(k_0) \rh^2_Q(k_0)e^{-\beta
    k_0} - 2\lambda \gamma(k_0) \rh(k_0)e^{-\beta k_0} - 2 \lambda_Q
  \rh_Q(k_0) \bigr].
\end{equation}
Its solution provides the spectral function
\begin{equation}
\label{rhoQsol}
   \rh_Q(k_0) = \frac{\lambda\gamma(k_0)+\lambda_Qe^{\beta
       k_0}}{\alpha_Q(k_0)}.
\end{equation}
The normalization condition can be satisfied if
\begin{equation}
  \lambda_Q = \frac{U_Q-\lambda X_1}{X_0},\qquad \lambda = \frac{s-
    \int_Q U_Q X_1/X_0}{\int_Q (X_2 - X_1^2/X_0)},
\end{equation}
where
\begin{equation}
  X_0 = \pintz{k_0}\frac{e^{\beta k_0}}{\alpha_Q(k_0)},\quad
  X_1 = \pintz{k_0}\frac{\gamma(k_0)}{\alpha_Q(k_0)},\quad
  X_2 = \pintz{k_0}\frac{\gamma^2(k_0)}{\alpha_Q(k_0)} e^{-\beta k_0}.
\end{equation}
Then the minimal $\eta_J$ reads
\begin{equation}
  \eta_J = \frac{\displaystyle \left(s - \int_Q
      \frac{U_QX_1}{X_0}\right)^2}{\displaystyle \int_Q \left(X_2 -
      \frac{X_1^2}{X_0} \right)^2} +  \int_Q \frac{U_Q^2}{X_0}.
\end{equation}

The concrete expressions for $X_i$ are potentially divergent. To
define still the result we have to introduce an energy cutoff
$\Lambda_0$ into the system. Then the qualitative behavior of the
$X_i$ quantities, assuming ${\cal J}^2(k_0)\sim k_0^a$:
\begin{equation}
  X_0 \sim \Lambda_0^{2-a} e^{\beta \Lambda_0},\qquad X_1 \sim
  \Lambda^{4-a},\qquad X_2\sim\mathrm{convergent}.
\end{equation}
Therefore removing the energy cutoff results in vanishing $X_1^n/X_0$
factors. This is exactly the mathematical appearance of the
qualitative analysis in the beginning of the subsection: the sum rule
constraints of \eqref{sumrule} can be easily satisfied by tuning only the
UV part of $\rh$.

When we remove the UV cutoff what remains is
\begin{equation}
  \eta_{J\mathrm{min}} = \frac{s^2}{\int_Q X_2}.
\end{equation}
Evaluating the integral in the denominator, for concreteness for shear
viscosity where we assume ${\cal J}\sim \k^2$, we obtain
\begin{equation}
  \frac{\eta}s\biggr|_{\mathrm{min}}\sim \frac{s}{N_Q L T^4},
\end{equation}
where $N_Q$ is the effective number of quantum channels. This formula
is valid for small $s$.

We remark here that the same line of thought leads to the conclusion
that also $\eta/\ep$ has a lower limit which is proportional to $\ep$
itself. Therefore $s$ has no singled out role in this analysis.

We can perform the same analysis for large $s$, too. This time we omit
the sum rule constraints, but we cannot omit ${\cal Z}$. Our strategy
will be to fix ${\cal Z} s$ and minimize ${\cal Z}\eta_J$. Therefore
we have to minimize the expression
\begin{equation}
  \min \left[ \int_Q\pintz{k_0} \alpha_Q(k_0)\rh_Q^2(k_0)e^{-\beta k_0} -
    2\lambda\left(\frac{{\cal Z}\ln{\cal Z}}{L^3} + \int_Q\pintz{k_0}
      \beta k_0 \rh_Q(k_0)e^{-\beta k_0}\right) \right],
\end{equation}
where
\begin{equation}
  {\cal Z} = 1+L^3\int_Q\pintz{k_0} \rh_Q(k_0)e^{-\beta k_0}.
\end{equation}
The minimum condition yields
\begin{equation}
  \rh_Q(k_0) = \lambda \frac{1+\ln{\cal Z}+\beta k_0}{\alpha_Q(k_0)}.
\end{equation}
If $\ln{\cal Z}\ll1$ we get back \eqref{rhoQsol}, this corresponds to
the small $s$ case. Here we follow the opposite limit and assume that
${\cal Z}$ dominates the above expression in the relevant $k_0\le T$
domain:
\begin{equation}
  \rh_Q(k_0) \approx \lambda \frac{\ln{\cal Z}}{\alpha_Q(k_0)}.
\end{equation}
With this density of states we can calculate
\begin{equation}
  {\cal Z} = 1 + \lambda Y_0 \ln{\cal Z} ,\qquad {\cal Z} s =
  \frac1{L^3} \left( {\cal Z}\ln{\cal Z} + \lambda Y_1 \ln{\cal
      Z}\right),
\end{equation}
where
\begin{equation}
  Y_0 = L^3 \int_Q\pintz{k_0} \frac{e^{-\beta k_0}}{\alpha_Q(k_0)},\qquad 
  Y_1 = L^3 \int_Q\pintz{k_0} \frac{\beta k_0 e^{-\beta
      k_0}}{\alpha_Q(k_0)}.
\end{equation}
For the case of the shear viscosity, assuming ${\cal J}_Q \sim \k^2$ we
obtain $Y_1\sim Y_0 \sim N_Q (LT)^4$. For ${\cal J}_Q=\k^2$ there is a
factor of $3$ between them, so we may assume that their values are of
similar order of magnitude in general. Then in the expression of
${\cal Z}s$ the first term is proportional to $(\ln{\cal Z})^2$, the
second only to $\ln{\cal Z}$. Therefore the second term can be
omitted, and we find
\begin{equation}
  {\cal Z} s \approx \frac{{\cal Z}\ln{\cal Z}}{L^3}
  \qquad\Rightarrow\qquad s = \frac{\ln{\cal Z}}{L^3}.
\end{equation}
Therefore in the large ${\cal Z}s$ regime $s$ and ${\cal Z}$ are not
independent. So if we fix a large value for ${\cal Z}s$ it means
fixing large value for separately $s$ and ${\cal Z}$. Large ${\cal Z}$
also means that the vacuum contribution can be omitted, and then
\begin{equation}
  \lambda = \frac{\cal Z}{Y_0 \ln{\cal Z}}.
\end{equation}
Finally for the transport coefficient we obtain
\begin{equation}
  \eta = \frac{\cal Z}{L^3 Y_0},
\end{equation}
therefore the $\eta/s$ ratio reads
\begin{equation}
  \frac{\eta}s\biggr|_{\mathrm{min}} = \frac{e^{L^3 s}}{Y_0L^3 s} \sim
  \frac{e^{L^3s}}{L^3 s} \frac 1{N_Q (LT)^4}.
\end{equation}

So we can summarize the small and large $s$ cases in a common formula
\begin{equation}
  \label{etaperesmin}
  \frac{\eta}s\biggr|_{\mathrm{min}} \sim \frac{{\cal F}(L^3 s)}{ N_Q (LT)^4 },
\end{equation}
where for small values ${\cal F}(x)\sim x$, for large values ${\cal
  F}(x) \sim e^x/x$.

The results has several interesting consequences. First of all we see
that if we distribute the entropy uniformly into $N_Q$ quantum
channels, we can reduce the $\eta/s$ ratio at any temperature to
zero. This corresponds to the construction of \cite{Cherman:2007fj} with large
number of particle species. In that model $L^3 s\sim \ln{N_Q}$ because
of the mixing entropy, which for large $N_Q$ leads to $1/\ln N_Q$
dependence for the shear viscosity to entropy ratio. In this
construction the entropy goes to infinity, while the shear viscosity
stays constant.

If we fix the number of dominant quantum channels then the $\eta/s$
ratio has a minimum for each fixed entropy, but this minimum depends
on the value of the entropy itself. The theoretical lower bound is
zero, which, however, can be reached only at zero entropy, ie. at zero
temperature.

If we have non-interacting quasiparticles, then the effective
interaction range is zero, $L\to0$ and the $\eta/s$ ratio is
infinity. In case of weak coupling the interaction range is related to
the cross section, or the inverse lifetime of the particle. A more
detailed description for the quasiparticle systems is done in the next
section.

\section{Quasiparticle systems}
\label{sec:QP}

Apart from the theoretical bound determined above, it is still a
question, how the $\eta/s$ ratio behaves in a real system. In this
Section we actualize the general formulae for quasiparticle systems,
and discuss the small width case.

For quasiparticles we use that the energy contribution from
non-interacting subsystems simply added in the free energy, so in
weakly interacting bosonic/fermionic system we can write
\begin{equation}
  f = T \pint4k \,\rh_{QP}(k) \, (\mp)\ln\left( 1\pm e^{-\beta k_0}\right).
\end{equation}
The entropy density can be calculated as $s=-\d f/\d T$.
\begin{equation}
\label{entropy}
  s = \sum\limits_Q \pint4k \left[ \frac{\beta k_0}{e^{\beta
        k_0} \pm 1} \pm \ln(1\pm e^{-\beta k_0}) \right] \rh_Q(k) \equiv
  \sum\limits_Q \pint4k {\cal X}(\beta k_0)\, \rh_Q(k).
\end{equation}

For concreteness we will calculate the shear viscosity coefficient
from now on. The viscosity to entropy density ratio reads
\begin{equation}
\label{etaperes}
  \frac{\eta}s = \frac{\displaystyle \frac\beta{15\cal Z}
    \sum_{\cal K} \pint4k \, \frac{(\k^2)^2}{k_0^2}\, e^{-\beta k_0}\,
    ({\cal J}_{\cal K}(k)\rh_{\cal K}(k))^2} {\displaystyle
    \sum\limits_Q \pint4k  {\cal X}(\beta k_0)\,\rh_Q(k)}.
\end{equation}
Formally it is similar to the perturbation theory motivated
expressions of \cite{Peshier:2005pp}. But here the logic is different,
we do not use propagating states, but instead energy eigenstates as
intermediate states.

To treat the problem analytically we will assume that ${\cal J}_{\cal
  K}(k)$ does not depend too strongly on the momentum on the relevant 
momentum regime, where, for example $\rh(k)$ exhibits a peak. We also
will assume that there is a single dominant quantum channel where the
largest contribution for entropy as well as the shear viscosity comes.
Therefore we analyze a reduced shear viscosity from the dominant
quantum channel as
\begin{equation}
\label{reducedeta}
  \bar\eta = \frac{\beta}{15}\,\pint4k \, \frac{(\k^2)^2}{k_0^2}\,
  e^{-\beta k_0}\, \rh_{\cal K}^2(k).
\end{equation}
The $\eta/s$ ratio then reads
\begin{equation}
\label{reducedetapers}
  \frac{\bar\eta}s = \frac{\displaystyle \frac\beta{15} \pint4k \,
    \frac{(\k^2)^2}{k_0^2} \, e^{-\beta k_0}\, \rh^2_{\cal K}(k)}
  {\displaystyle \pint4k  {\cal X}(\beta k_0)\,\rh_Q(k)}.
\end{equation}
We hope that $\bar\eta$ characterizes well the true $\eta/s$, so we will
use this quantity to assess the importance of the off-shell effects.

\subsection{Small width approximation}

If the integral is dominated by a Dirac-delta-like peak then near the
peak region we may approximate
\begin{equation}
  \label{QPappr}
  \rh_{\cal K}(k)\biggr|_{k_0\approx\ep_k}
  \approx\frac{2\Gamma}{(k_0-\ep_k)^2+\Gamma^2}. 
\end{equation}
In relativistically invariant systems $\ep_k^2=k^2+m^2$, and
approximately the same is true at finite temperatures with thermal
masses.

The product of two functions can be approximated by
\begin{equation}
  \rh_{\cal K}^2(k) \approx \rh_{\cal K}(k_0=\ep_k) \rh(k) \approx
  \frac2\Gamma\,2\pi\delta(k_0-\ep_k).
\end{equation}
By replacing back this expression into \eqref{reducedeta} we find
\begin{equation}
  \bar\eta_{QP} = \frac1{15\pi\Gamma T} \int\limits_0^\infty
  dk\,\frac{k^6}{\ep_k^2} e^{-\ep_k/T},
\end{equation}
where the $QP$ subscript refers to the small-width quasiparticle case.
Depending on the dispersion relation we can obtain results like
\begin{equation}
  \bar \eta_{QP} = \left\{
    \begin{array}[c]{ll}
      \displaystyle \rule[-1em]{0em}{1em} \frac{8}{5\pi^2}
      \frac{T^4}\Gamma,\qquad&\mathrm{if}\; \ep_k=k\cr 
      \displaystyle \rule[-1em]{0em}{1em} \frac1{\sqrt{2\pi}}
      \frac{m^{3/2} T^{5/2}}\Gamma\, e^{-m/T},\qquad&\mathrm{if}\;
      \ep_k = m + \frac{k^2}{2m}\cr
      \displaystyle \frac{4}{15\sqrt{2\pi}} \frac{m^{7/2} T^{1/2}}\Gamma
      ,\qquad&\mathrm{if}\; \ep_k =\frac{k^2}{2m}\cr
    \end{array}\right.
\end{equation}
The first case corresponds the high temperature case, the second the
low temperature relativistic case, the third a low temperature
nonrelativistic gas case (Chapman-Enskog formula).

The free energy in the quasiparticle case \eqref{QPappr} reads after
partial integration: 
\begin{equation}
  f_{QP} = -\frac1{6\pi^2} \int\limits_{E_0}^\infty\! dE
  \,k(E)^3 n(E),
\end{equation}
where $n_\pm(E) =(e^{\beta E}\pm1)^{-1}$ Bose-Einstein or Fermi-Dirac
particle number density. For the entropy we obtain
\begin{equation}
  s = \frac{\beta}{6\pi^2} \int\limits_{E_0}^\infty dE \, n_\pm(E)
  \frac{d}{dE} (Ek^3(E)).
\end{equation}
In the different dispersion relation case we get:
\begin{equation}
  s_{QP} = \left\{
    \begin{array}[c]{ll}
      \displaystyle \rule[-1em]{0em}{1em} {\cal A}_\pm
      \frac{4T^3\pi^2}{90},\qquad & \mathrm{if}\; \ep_k=k\;\cr 
      \displaystyle \rule[-1em]{0em}{1em} \frac{T^{1/2}
        m^{5/2}}{(2\pi)^{3/2}} e^{-\beta m},\qquad&\mathrm{if}\; \ep_k
      = m+\frac{k^2}{2m}\cr
      \displaystyle \frac{(mT)^{3/2}}{2(2\pi)^{3/2}},\qquad&\mathrm{if}\;
      \ep_k  =\frac{k^2}{2m},\cr
    \end{array}\right.,
\end{equation}
where ${\cal A}_\pm =(1,8/7)$. Then the ratio reads
\begin{equation}
  \frac{\bar\eta_{QP}}{s_{QP}} = \left\{
    \begin{array}[c]{ll}
      \displaystyle \rule[-1em]{0em}{1em} \frac{36}{{\cal A}_\pm
        \pi^4} \frac T\Gamma,\qquad& \mathrm{if}\; \ep_k=k\;\cr 
      \displaystyle \rule[-1em]{0em}{1em} 2\pi\,\frac{T^2}{\Gamma
        m},\qquad &\mathrm{if}\; \ep_k = m+\frac{k^2}{2m}\cr
      \displaystyle \frac{16\pi}{15}\, \frac{m^2}{\Gamma
          T},\qquad&\mathrm{if}\; \ep_k  =\frac{k^2}{2m}\cr
    \end{array}\right.
\end{equation}

In the first, $m\ll T$ case the width must determined by the
temperature which is the only scale in the system: $\Gamma\sim T$ by
dimensional reasons. Therefore in the first case we obtain temperature
independent ratio.

In the second, low temperature massive case we have to take into account
that the quasiparticles are stable at zero temperature, and the first
scattering states which are responsible for the decay have to be
excited thermally. Therefore we expect that $\Gamma \sim e^{-M/T}$,
where $M$ is the mass of the excited state. In this case the ratio
grows like $T^2 e^{M/T}$ at small temperature. 

In the third case the width can be proportional to $T$, but also in
this case the ratio will grow like $1/T^2$.

So we can see that in the quasiparticle case the $\eta/s$ ratio is at
most constant, but in all massive cases it is bounded from below. The
lower bound of this ratio is probably provided by the AdS/CFT results.

The inverse proportionality to $\Gamma$ also suggests that in
environments where the correlation lengths, in particular the
quasiparticle lifetimes grow, the $\eta/s$ ratio will decrease. This
is the case for second order phase transitions.

\section{Relevance of the off-shell effects in different physical situations}
\label{sec:offshell}

In the field theoretical systems besides the (quasi)particles one
should count with a multiparticle contribution which appears as a
continuum in the energy density of states. The simple picture
suggested by the small width quasiparticle approximation of the previous
subsection may break down more or less in case of strong continuum.

In this section we consider three physical situations where the
presence of off-shell effects can significantly modify the small width
quasiparticle picture. The first is the effect of the wave function
renormalization, the second is the case of large thermal width, the
third is the case of a theory with zero mass excitations.

\subsection{Wave function renormalization}

As a first example we take a system where besides the quasiparticle
peak there is a separate continuum. The normalization requirement
\eqref{sumrule} tells us that in the full case the peak should be smaller
than in the quasiparticle approximation: the factor we have to apply
is the wave function renormalization. At small enough temperatures we
will see only the quasiparticle peak, then the $\eta/s$ ratio is
simply rescaled by the wave function renormalization factor.

To demonstrate the effect we take a toy model, where the full spectral
density is taken as
\begin{equation}
  \rh(k) = Z \rh_{QP}(k) + 2\pi \frac{1-Z}{E_2-E_1} \Theta(E_2>k_0>E_1),
\end{equation}
ie. we take a step-function-like spectral density in the continuum
regime, which means that in this toy model we do not care about the
details of the continuum, only an effective height is taken into
account. Here we take the relativistic form $E_{1,2}^2= \k^2 +
M_{1,2}^2$, while $\rh_{QP}$ has the form of \eqref{QPappr}. The
proportionality factors are taken into account to satisfy the sum rule
\begin{equation}
  \pintz{k_0} \rh(k_0) = \pintz{k_0} \rh_{QP}(k_0) =1.
\end{equation}
We will assume the hierarchy $m\ll M_{1,2}$ and $T< M_1$. In this
limit in the square $\rh^2$ we can neglect the cross terms, and we can
use nonrelativistic approximation in the $E_{1,2}$ energies. What we
obtain in the leading order
\begin{equation}
  \bar \eta = Z^2 \bar\eta_{QP} + \frac1{5\sqrt{2\pi}}\,
  \frac{(1-Z)^2 T^{5/2}M_1^{5/2}}{\Delta M^2}\, e^{-M_1/T},
\end{equation}
where $\Delta M=M_2-M_1$. With similar assumptions the entropy density
reads:
\begin{equation}
  s = Zs_{QP} + \frac1{(2\pi)^{3/2}}\frac{(1-Z) T^{3/2} M_1^{5/2}}
  {\Delta M}\, e^{-M_1/T}.
\end{equation}

If the temperature is lower than the threshold, then the radiative
corrections for both $\bar\eta$ and $s$ are suppressed. Then
approximately we obtain
\begin{equation}
  \frac{\bar\eta}s = Z\, \frac{\bar\eta_{QP}}{s_{QP}}.
\end{equation}
That means that even though the continuum contribution cannot be seen,
its effect is measurable in the $\eta/s$ ratio.

The value of $Z$ is determined by the relative contribution of the
continuum in the sum rule as compared to the quasiparticle peak. In
weak coupling cases this ratio is suppressed by (powers of) the
coupling constant, so $Z\approx 1$ and so the effect on $\eta/s$ is
small. However, in strongly coupled theories the continuum becomes
more and more important, and, correspondingly, the wave function
decreases. As $Z\to0$ the continuum contribution slowly takes over and
it will dominate the $\eta/s$ ratio. In this simple case the
theoretical $Z\to0$ limit the continuum yields $\bar\eta/s \sim T/\Delta
M$. This demonstrates a mathematical example where in a constructed
system the $\eta/s$ ratio can go to zero. Still, we have to remark
that there does not exist a physical system where this type of
continuum would be manifested. But it emphasizes the importance of the
continuum part of the spectral function in the transport. Two more realistic
cases will be studied in the next subsections.

\subsection{Large width case}

In lots of realistic examples there is a quasiparticle peak, but its
width is large, comparable to the mass scales of the system, cf. for
example \cite{Peshier:2005pp}. At finite temperature strongly coupled
theories, moreover, the quasiparticle peak can merge with the
continuum, forming a broad, slowly varying spectral function, as it
can be seen for example in the 2PI simulations
\cite{Arrizabalaga:2006hj, Jakovac:2006gi, Arrizabalaga:2007zz}. In
this case the ``width'' of the peak has nothing to do with the
quasiparticles. In fact there are no real particles in the plasma, the
excitations decay before a particle is formed. Correspondingly, as the
width grows, the quasiparticle picture becomes less and less good
approximation.

To treat this effect numerically in the case of the transport
coefficients, we take a spectral function which could describe a
spectral function of a quantum channel at finite temperature, and
which is of a Breit-Wigner-type form
\begin{equation}
  \rh_{BW}(k) = \frac1{\cal N} \frac{4\gamma_k^2 k_0}{(k_0^2-E_k^2)^2 +
    \gamma_k^4}.
\end{equation}
The ${\cal N}$ prefactor is necessary for normalization, is ${\cal
  N}=(\pi + 2\arctan(E_k^2/\gamma_k^2))/(2\pi)$. In the above formula the
dispersion relation and the momentum dependence of the $\gamma$
parameter is arbitrary, but we assume spatial rotational invariance.
In the weak coupling limit the width corresponds to
$\Gamma=\gamma^2/(2E_k)$. In the high temperature field theories we
expect $\gamma\sim g^2 T$. In strongly coupled case $\gamma \gg T$
is also possible.

In the $\bar\eta/s$ formula we rescale by the temperature, so we
choose $T=1$. Then we have to calculate
\begin{equation}
  \frac{\bar\eta}s = \frac{\displaystyle \int\limits_0^\infty dk\,k^6
    \int \limits_0^\infty dk_0\,e^{-k_0}\,(\rh_{BW}/k_0)^2 }
  {\displaystyle 15\int\limits_0^\infty dk\,k^2 \int \limits_0^\infty
    dk_0\,\left(k_0 (e^{k_0}-1)^{-1} - \ln (1-e^{-k_0})\right) \rh_{BW}},
\end{equation}
where we have chosen, for the sake of concreteness, the minus sign in
${\cal X}$.

\begin{figure}[htbp]
  \centering
  \includegraphics[height=5cm]{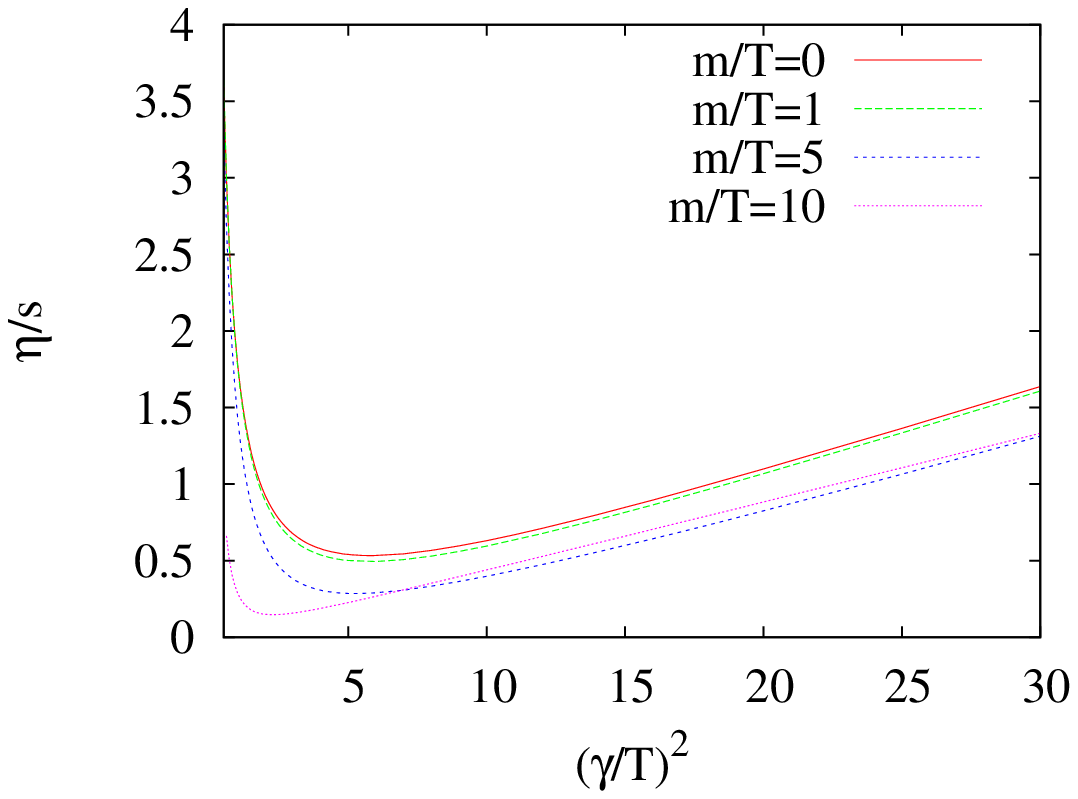}
  \includegraphics[height=5cm]{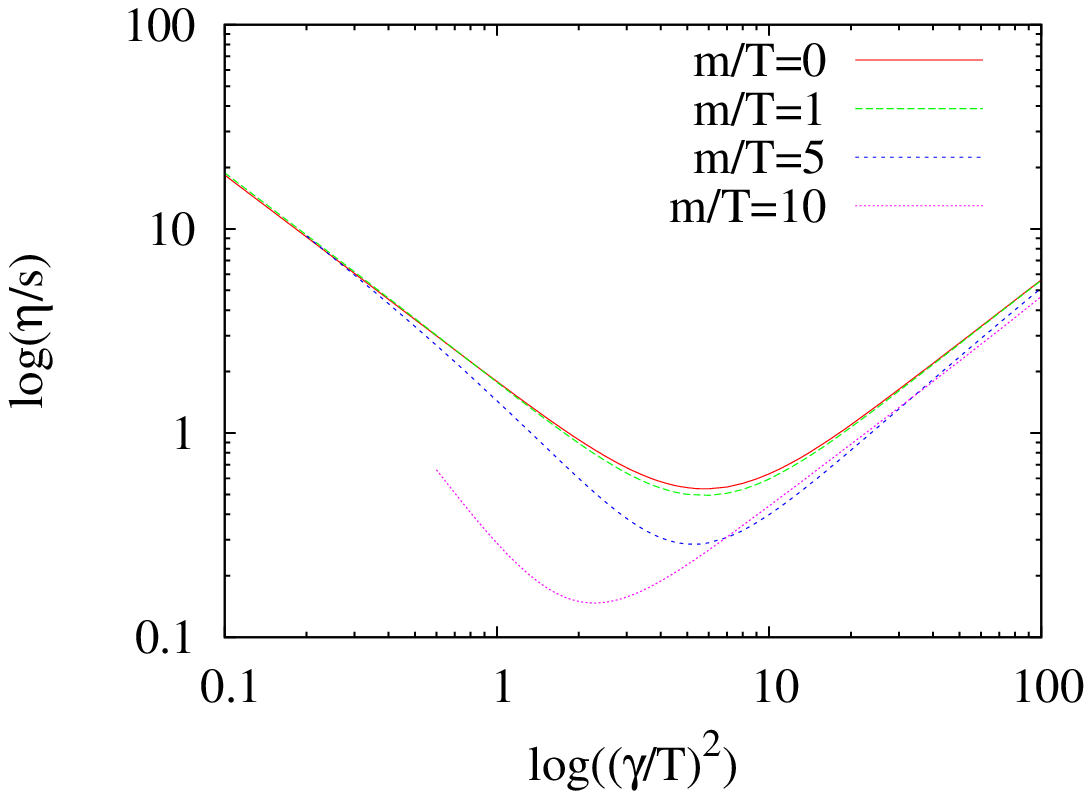}
  \caption{The shear viscosity to entropy ratio as a function of the
    imaginary self energy.}
  \label{fig:etaperes1}
\end{figure}

For a particular choice of $\gamma_k=\gamma =\mathrm{const}$ and
$E_k=\sqrt{k^2+m^2}$ we find the plot of the ratio in
Fig.~\ref{fig:etaperes1}.
This figure shows how the small width quasiparticle picture which
yields in this case $\sim T^2/\gamma^2$ breaks down for larger width
case, where we find $\sim \gamma^2/T^2$ behavior. Looking at these
figures as function of temperature at fixed $\gamma$ we see that at
high temperatures the $\eta/s$ ratio increases with increasing
temperature, while at low temperature it decreases with increasing
temperature. While the former is characteristic for gases, the latter
is the behavior of the fluids. So in this approximation we can give
an account for the $\eta/s$ ratio in the fluid-gas crossover.

How can we understand mathematically this behavior? The small width
regime corresponds to the quasiparticle approximation above, so we have
to understand the large $\gamma$ regime, so examine the $\gamma\gg T$
case. Because of the exponential suppression for $k_0$, the value of
$k_0$ cannot grow larger than $T$. The peak region, however, is in the
vicinity of $|k_0^2 - E_k^2| \sim \gamma^2$. Therefore $k_0$ cannot
play a role in saturating the integrand (unless we are in the
quasiparticle regime $\gamma\ll m$), and so we can neglect $k_0$
here. Then the dominant contribution should come from $E_k\sim\gamma$,
since there is no exponential suppression for $k$. As a result $k_0\ll
k$ and the $k_0$ and $k$ integrations decouple. For $\gamma\gg m$ case
$E_k\sim\gamma\gg m$ means that we can use ultrarelativistic
dispersion relation. Then for the numerator we have
\begin{equation}
  \bar\eta\to \int \limits_0^\infty dk_0\,e^{-k_0} \int\limits_0^\infty
    dk\,k^6\, \frac1{{\cal N}^2} \frac{16 \gamma^4}{(k^4
      +\gamma^4)^2} \sim \gamma^3,
\end{equation}
and for the denominator we obtain
\begin{equation}
  s\to \int \limits_0^\infty dk_0\,k_0 \left(k_0 (e^{k_0}-1)^{-1} - \ln
    (1-e^{-k_0})\right) \int\limits_0^\infty dk\,k^2 \frac1{\cal N}
  \frac{4\gamma^2}{k^4+\gamma_k^4} \sim \gamma.
\end{equation}
Therefore we expect that the ratio grows like $\gamma^2/T^2$. In the
opposite case, when $\gamma\ll m$, but the decoupling of the integrals
is still true, then we can neglect the $\gamma$ factor in the
denominator and we obtain
\begin{equation}
  \bar\eta\to \int \limits_0^\infty dk_0\,e^{-k_0} \int\limits_0^\infty
    dk\,k^6\, \frac1{{\cal N}^2} \frac{16 \gamma^4}{((k^2+m^2)^2} \sim
    \frac{\gamma^4}{m^2},
\end{equation}
and 
\begin{equation}
  s\to \int \limits_0^\infty dk_0\,k_0 \left(k_0 (e^{k_0}-1)^{-1} - \ln
    (1-e^{-k_0})\right) \int\limits_0^\infty dk\,k^2 \frac1{\cal N}
  \frac{4\gamma^2}{(k^2+m^2)^2} \sim \frac{\gamma^2}{m}.
\end{equation}
Therefore the ratio behaves as $\gamma^2/m$, so the trend of large
$\gamma$ regime continues.

To have a feeling of when does the small-width description start to
dominate, we remark that the quasiparticle contribution is at one hand
suppressed by $e^{-E_k}$, but, on the other hand, the denominator can
be small, too. Decoupling of the integrals will not be true if the
denominator can compensate the exponential suppression. In the small
mass case, when also $\gamma\gg m$ is true, the quasiparticle regime
gives $e^{-k}/\gamma^8$ with $k\sim \gamma$, as opposed to the
$1/4\gamma^8$ factorized case. So on-shell effects will become strong
if $\gamma\lsim 1.4$. In the large mass case the on-shell contribution
is $e^{-E_k}/\gamma^8$ with $E_k\sim m$, the off-shell effects yield
$1/m^8$. Then the borderline is at $\gamma\sim m e^{-m/8}$. In the
hypothetical $m\to\infty$ case, therefore, there is no quasiparticle
regime, the linear $\gamma^2$ trend will continue until $\gamma=0$,
where the ratio is zero.

\subsection{Low temperature systems with zero mass excitation}

If we wonder whether we can reach in a real system zero value for
$\eta/s$, with finite number of quantum channels, we must reach,
according to \eqref{etaperesmin}, zero entropy. In a real system can
happen only at zero temperature. In the small-width case, as we have
seen, the $\eta/s$ ratio diverges in the massive case, and, formally,
stay constant in the zero mass case.

The zero mass systems, however, need a more detailed study, since
there are never classical Dirac-delta-like spectral functions. To see
this, we recall that if we take a quantum field theoretical system
with a stable particle with mass $m$ which interacts with another
particle with mass $M$ then the spectral function of this system will
contain a Dirac-delta peak at $k_0=m$ and a continuum starting at the
threshold $m+M$. The imaginary part of the self energy, because of
kinematical reasons, is proportional to $\sqrt{\lambda(p^2,m^2,M^2)}$,
where $\lambda(x,y,z)=(x-y-z)^2-4y^2z^2$. For a finite $M$ this leads
to a square-root behavior for the complete spectral function,
$\rh\sim \sqrt{p^2-(m+M)^2}$ near the threshold.

If the massive particle interacts with a zero mass particle, all this
means that there is no gap between the quasiparticle peak and the
continuum. Moreover, in this case $\sqrt{\lambda(p^2,m^2,0)}=p^2-m^2$,
which is the same as the tree level part of the real part. This has
the consequence that near the threshold the spectral function is
divergent $\rh \sim (p^2-m^2)^{-1}$. There are logarithmic corrections
to this behavior; for example in QED at one loop level in Lorenz
gauge \cite{PeshkinSchroeder}
\begin{equation}
  \rh(p^2\approx m^2) = \frac{\displaystyle 2\Theta(p^2-m^2)
    \frac\alpha\pi}{\displaystyle (\slash\hspace{-0.5em}p -
    m)\left(\left(1+\frac\alpha\pi \ln \left|\frac{p^2-m^2}{\mu^2}
        \right| \right)^2 + \frac{\alpha^2}{\pi^2}\right)}. 
\end{equation}
The leading logarithmic corrections can be resummed by renormalization
group or by Bloch-Nordsieck construction \cite{BogShir}, the result is
the modification of the threshold behavior
\begin{equation}
  \rh(p^2\approx m^2) = \frac{2\Theta(p^2-m^2)(\slash\hspace{-0.5em}p + m)
    m^{-2\beta}\sin\alpha} {(p^2-m^2)^{1-\beta}},\qquad \beta =
  -\frac{\alpha}\pi,\; \alpha=\frac{e^2}{4\pi}.
\end{equation}
The main features, however, still survived: there is no gap between
the quasiparticle peak and the continuum, and there is a rising
spectral function as we approach the mass shell.

This behavior can be studied numerically, too. In case of QCD one can
use Monte Carlo data to reconstruct the spectral function
\cite{Biro:2006iy}, which exhibits qualitatively the same behavior as
in the QED case.

According to the above analysis in \emph{all} systems containing zero
mass particles, the energy density of states cannot contain a zero
width quasiparticle. The would-be quasiparticle continuously radiates
long-living soft zero mass excitations, and continuously interacts
with them. The spectral function is not Lorentzian, it has no
``width'' in the usual sense, and it does not describe a decay,
too. At larger temperature we will still observe a quasiparticle,
although the mechanism how it is formed is far from trivial
\cite{Blaizot:1996az}. At low temperatures, however, we are in a
non-quasiparticle system, where the effect of the threshold behavior
cannot be neglected.

At small temperatures we are in an almost Lorentz invariant
system. Because of the volume normalization we can write $\rh(k) = k_0
\bar\rh(k^2)$, where $\bar \rh$ depends only on the Lorentz invariant
$k^2$ form. Then we write (cf. \eqref{reducedetapers})
\begin{equation}
  \label{smT}
  \frac{\bar\eta}s = \frac{\displaystyle \beta \pint4k \, (\k^2)^2
    e^{-\beta k_0}\, {\bar \rh}^2(k)} {\displaystyle 15\pint4k k_0{\cal
      X}(\beta k_0) e^{-\beta k_0}\,\bar\rh(k)}.
\end{equation}
The exponential factor forces the system to have the possible lowest
energy values, ie. near the threshold. 

First let us assume that the lowest lying threshold is at $k^2=M^2$
with finite mass. This is conceivable if the zero mass particle is not
asymptotic state, like in low energy QCD. Near the colorless bound
states, which have color multipole moments, the gluons can still
exist, and the effective gluon cloud can result in a continuous
threshold behavior. In this case we can use $\beta k_0>\beta M\gg1$, and so
${\cal X}(\beta k_0) \approx \beta k_0$. We change into 4D
polar-coordinates ($k_0=k\cosh\eta,\, k_z = k\sinh\eta
\cos\theta,\,k_y=k\sinh\eta\sin\theta\sin\phi,\, k_x=k\sinh\eta
\sin\theta\cos\phi$), taking into account that $k_0>0$, and the 3D
rotational invariance:
\begin{equation}
  \pint4k \Theta(k_0) \to \frac1{4\pi^3} \int\limits_0^\infty dk k^3
  \int\limits_0^\infty d\eta \sinh^2\eta.
\end{equation}
We use the integral formula
\begin{equation}
  \int\limits_0^\infty d\eta \sinh^n\eta e^{-z\cosh\eta} =
  \frac{K_{n/2}(z) \Gamma\left(\frac{n+1}2\right)}
  {\sqrt\pi\,(z/2)^{n/2}}
\end{equation}
($K_n$ is the modified Bessel function of the second kind), and the
asymptotic form $K_n(x) =\sqrt{\pi/(2x)} e^{-x}$ to arrive at:
\begin{equation}
  \frac{\bar\eta}s = \frac{T^2\int\limits_0^\infty dk\,k^{7/2}\,
    e^{-\beta k} \bar\rh^2(k)}{\int\limits_0^\infty dk\,k^{7/2}\,
    e^{-\beta k} \bar\rh(k)}
\end{equation}
Near the threshold we power expand the spectral function like
\begin{equation}
  \bar\rh(k\approx M) = {\cal C} \Theta(k-M) \,(k^2-M^2)^w.
\end{equation}
And then find
\begin{equation}
  \frac{\bar\eta}s = \frac{\Gamma(1+2w)}{\Gamma(1+w)} \,{\cal
    C}(2M)^w\, T^{2+w}.
\end{equation}
This result cleanly shows that if there exist a system described
above, it must have a vanishing $\eta/s$ ratio in the $T\to0$ limit. 

If the zero mass particle is an asymptotic state, then the lowest
lying energy eigenvalues belong to the quantum channel of the massless
particle. Since it interacts with itself, the spectral function has no
gap here, too. A special class is the conform theories, another the
weakly interacting gauge bosons, like the photon gas, where the
photon-photon scattering is mediated by virtual electron loop. In the
massless case the above analysis goes through without modification and
it yields
\begin{equation}
  \frac{\bar\eta}s \sim {\cal C} T^{2(1+w)}.
\end{equation}
This result could be obtained by purely dimensional analysis, since
the dimension of ${\cal C}$ is $-2(1+w)$. The numerator of \eqref{smT}
contains $\rh^2$, the denominator only $\rh$, so there remains a
factor ${\cal C}$ in the ratio. Since the ratio is dimensionless,
something has to compensate this factor. In the massless case, only
the temperature can do that: thus we obtain $\sim T^{2(1+w)}$
dependence.

In conformal field theories without anomalous dimensions ${\cal C}$ is
dimensionless and $w=-1$. That predicts a temperature-independent
$\eta/s$ ratio. In other cases, ie. either in a non-conformal field
theory or in a conformal field theory with anomalous dimension $w$ can
differ from $-1$, then we again observe a vanishing $\eta/s$ ratio in
the zero temperature limit.

\section{Conclusion}
\label{sec:concl}

In this paper we examined the shear viscosity to entropy density ratio
using exact representation through the density of states or energy
spectral functions. We examined what can be said purely mathematically
about the ratio, assuming some physical conditions (sum rules) for the
spectral functions, and keeping the entropy density constant. We
concluded that the $\eta/s$ ratio has no lower bound in a most generic
class of physical systems. To understand this statement qualitatively
we recall that $\eta\sim \rh^2$ and $s\sim\rh$ if the temperature is
low enough. Therefore if the density of states exhibits large peaks
(quasiparticle systems) then $\rh^2\gg\rh$ and so the $\eta/s$ ratio
is large; if $\rh$ is small everywhere then $\rh^2\ll\rh$ and the
ratio is small. However, if we fix the entropy density then there is a
minimum, where $\eta\sim s^2$. This means that $\eta/s$ can go to zero
(in a fixed system) only at zero entropy density, ie. at zero
temperature.

Although mathematically the situation is clear, it is hard to show
physical systems where a small $\eta/s$ ratio can occur. First of all
if it consists of quasiparticle at all, they must not have small
width, since the small width approximation excludes small $\eta/s$
ratio. Candidates for such a system are quasiparticle systems with
strong continuum (small wave function renormalization constant), high
temperature strongly interacting systems or low temperature systems
with zero mass excitations. In the first two cases it is possible to go
below $1/4\pi$, but we can never reach zero value. In the third case
if the threshold behavior is $(k^2-M^2)^w$ then the $\eta/s$ ratio can
reach zero as $T^{2(1+w)}$ (in $M=0$ case) or $T^{2+w}$ (massive
case). So if $w>-1$ then we will find a vanishing ratio in the $T\to
0$ limit.

As far as the heavy ion experiments are concerned, the QCD at finite
temperature and density represents a finite entropy density
system. Therefore there is a lower bound for the $\eta/s$ ratio, but
its value is not necessarily $1/4\pi$. However, the relevant scales
are set either by $\Lambda_{QCD}$ or the temperature (since the
couplings are of order one), and these two scales are again
similar. Therefore we are close to a one-scale system, where the
predictions of the conformal theory may apply.

\section*{Acknowledgment}

The author acknowledges useful discussions with C. Greiner, A. Patkos,
A. Peshier, Zs. Szep, G. Zarand and A. Zawadowski, and geting useful
hints from M. Gyulassy. This work was supported by the Hungarian
Science Fund (OTKA) K68108.

\bibliography{hsratio}

\providecommand{\noopsort}[1]{}\providecommand{\singleletter}[1]{#1}%
\begin{thebibliography}{39}
\expandafter\ifx\csname natexlab\endcsname\relax\def\natexlab#1{#1}\fi
\expandafter\ifx\csname bibnamefont\endcsname\relax
  \def\bibnamefont#1{#1}\fi
\expandafter\ifx\csname bibfnamefont\endcsname\relax
  \def\bibfnamefont#1{#1}\fi
\expandafter\ifx\csname citenamefont\endcsname\relax
  \def\citenamefont#1{#1}\fi
\expandafter\ifx\csname url\endcsname\relax
  \def\url#1{\texttt{#1}}\fi
\expandafter\ifx\csname urlprefix\endcsname\relax\def\urlprefix{URL }\fi
\providecommand{\bibinfo}[2]{#2}
\providecommand{\eprint}[2][]{\url{#2}}

\bibitem[{\citenamefont{Adler et~al.}(2003)}]{Adler:2003kt}
\bibinfo{author}{\bibfnamefont{S.~S.} \bibnamefont{Adler}} \bibnamefont{et~al.}
  (\bibinfo{collaboration}{PHENIX}), \bibinfo{journal}{Phys. Rev. Lett.}
  \textbf{\bibinfo{volume}{91}}, \bibinfo{pages}{182301}
  (\bibinfo{year}{2003}), \eprint{nucl-ex/0305013}.

\bibitem[{\citenamefont{Adams et~al.}(2004)}]{Adams:2003am}
\bibinfo{author}{\bibfnamefont{J.}~\bibnamefont{Adams}} \bibnamefont{et~al.}
  (\bibinfo{collaboration}{STAR}), \bibinfo{journal}{Phys. Rev. Lett.}
  \textbf{\bibinfo{volume}{92}}, \bibinfo{pages}{052302}
  (\bibinfo{year}{2004}), \eprint{nucl-ex/0306007}.

\bibitem[{\citenamefont{Shuryak}(2004)}]{Shuryak:2003xe}
\bibinfo{author}{\bibfnamefont{E.}~\bibnamefont{Shuryak}},
  \bibinfo{journal}{Prog. Part. Nucl. Phys.} \textbf{\bibinfo{volume}{53}},
  \bibinfo{pages}{273} (\bibinfo{year}{2004}), \eprint{hep-ph/0312227}.

\bibitem[{\citenamefont{Teaney}(2003)}]{Teaney:2003kp}
\bibinfo{author}{\bibfnamefont{D.}~\bibnamefont{Teaney}},
  \bibinfo{journal}{Phys. Rev.} \textbf{\bibinfo{volume}{C68}},
  \bibinfo{pages}{034913} (\bibinfo{year}{2003}), \eprint{nucl-th/0301099}.

\bibitem[{\citenamefont{Romatschke and Romatschke}(2007)}]{Romatschke:2007mq}
\bibinfo{author}{\bibfnamefont{P.}~\bibnamefont{Romatschke}} \bibnamefont{and}
  \bibinfo{author}{\bibfnamefont{U.}~\bibnamefont{Romatschke}},
  \bibinfo{journal}{Phys. Rev. Lett.} \textbf{\bibinfo{volume}{99}},
  \bibinfo{pages}{172301} (\bibinfo{year}{2007}), \eprint{0706.1522}.

\bibitem[{\citenamefont{Heinz and Song}(2008)}]{Heinz:2008qm}
\bibinfo{author}{\bibfnamefont{U.~W.} \bibnamefont{Heinz}} \bibnamefont{and}
  \bibinfo{author}{\bibfnamefont{H.}~\bibnamefont{Song}}, \bibinfo{journal}{J.
  Phys.} \textbf{\bibinfo{volume}{G35}}, \bibinfo{pages}{104126}
  (\bibinfo{year}{2008}), \eprint{0806.0352}.

\bibitem[{\citenamefont{Jeon}(1995)}]{Jeon:1994if}
\bibinfo{author}{\bibfnamefont{S.}~\bibnamefont{Jeon}}, \bibinfo{journal}{Phys.
  Rev.} \textbf{\bibinfo{volume}{D52}}, \bibinfo{pages}{3591}
  (\bibinfo{year}{1995}), \eprint{hep-ph/9409250}.

\bibitem[{\citenamefont{Wang and Heinz}(2003)}]{Wang:2002nba}
\bibinfo{author}{\bibfnamefont{E.}~\bibnamefont{Wang}} \bibnamefont{and}
  \bibinfo{author}{\bibfnamefont{U.~W.} \bibnamefont{Heinz}},
  \bibinfo{journal}{Phys. Rev.} \textbf{\bibinfo{volume}{D67}},
  \bibinfo{pages}{025022} (\bibinfo{year}{2003}), \eprint{hep-th/0201116}.

\bibitem[{\citenamefont{Aarts and Martinez~Resco}(2005)}]{Aarts:2005vc}
\bibinfo{author}{\bibfnamefont{G.}~\bibnamefont{Aarts}} \bibnamefont{and}
  \bibinfo{author}{\bibfnamefont{J.~M.} \bibnamefont{Martinez~Resco}},
  \bibinfo{journal}{JHEP} \textbf{\bibinfo{volume}{03}}, \bibinfo{pages}{074}
  (\bibinfo{year}{2005}), \eprint{hep-ph/0503161}.

\bibitem[{\citenamefont{Gagnon and Jeon}(2007)}]{Gagnon:2007qt}
\bibinfo{author}{\bibfnamefont{J.-S.} \bibnamefont{Gagnon}} \bibnamefont{and}
  \bibinfo{author}{\bibfnamefont{S.}~\bibnamefont{Jeon}},
  \bibinfo{journal}{Phys. Rev.} \textbf{\bibinfo{volume}{D76}},
  \bibinfo{pages}{105019} (\bibinfo{year}{2007}), \eprint{0708.1631}.

\bibitem[{\citenamefont{Huot et~al.}(2007)\citenamefont{Huot, Jeon, and
  Moore}}]{Huot:2006ys}
\bibinfo{author}{\bibfnamefont{S.~C.} \bibnamefont{Huot}},
  \bibinfo{author}{\bibfnamefont{S.}~\bibnamefont{Jeon}}, \bibnamefont{and}
  \bibinfo{author}{\bibfnamefont{G.~D.} \bibnamefont{Moore}},
  \bibinfo{journal}{Phys. Rev. Lett.} \textbf{\bibinfo{volume}{98}},
  \bibinfo{pages}{172303} (\bibinfo{year}{2007}), \eprint{hep-ph/0608062}.

\bibitem[{\citenamefont{Arnold et~al.}(2003)\citenamefont{Arnold, Moore, and
  Yaffe}}]{Arnold:2002zm}
\bibinfo{author}{\bibfnamefont{P.}~\bibnamefont{Arnold}},
  \bibinfo{author}{\bibfnamefont{G.~D.} \bibnamefont{Moore}}, \bibnamefont{and}
  \bibinfo{author}{\bibfnamefont{L.~G.} \bibnamefont{Yaffe}},
  \bibinfo{journal}{JHEP} \textbf{\bibinfo{volume}{01}}, \bibinfo{pages}{030}
  (\bibinfo{year}{2003}), \eprint{hep-ph/0209353}.

\bibitem[{\citenamefont{Jeon and Yaffe}(1996)}]{Jeon:1995zm}
\bibinfo{author}{\bibfnamefont{S.}~\bibnamefont{Jeon}} \bibnamefont{and}
  \bibinfo{author}{\bibfnamefont{L.~G.} \bibnamefont{Yaffe}},
  \bibinfo{journal}{Phys. Rev.} \textbf{\bibinfo{volume}{D53}},
  \bibinfo{pages}{5799} (\bibinfo{year}{1996}), \eprint{hep-ph/9512263}.

\bibitem[{\citenamefont{Xu and Greiner}(2008)}]{Xu:2007ns}
\bibinfo{author}{\bibfnamefont{Z.}~\bibnamefont{Xu}} \bibnamefont{and}
  \bibinfo{author}{\bibfnamefont{C.}~\bibnamefont{Greiner}},
  \bibinfo{journal}{Phys. Rev. Lett.} \textbf{\bibinfo{volume}{100}},
  \bibinfo{pages}{172301} (\bibinfo{year}{2008}), \eprint{0710.5719}.

\bibitem[{\citenamefont{Xu et~al.}(2008)\citenamefont{Xu, Greiner, and
  Stocker}}]{Xu:2007jv}
\bibinfo{author}{\bibfnamefont{Z.}~\bibnamefont{Xu}},
  \bibinfo{author}{\bibfnamefont{C.}~\bibnamefont{Greiner}}, \bibnamefont{and}
  \bibinfo{author}{\bibfnamefont{H.}~\bibnamefont{Stocker}},
  \bibinfo{journal}{Phys. Rev. Lett.} \textbf{\bibinfo{volume}{101}},
  \bibinfo{pages}{082302} (\bibinfo{year}{2008}), \eprint{0711.0961}.

\bibitem[{\citenamefont{Petreczky}(2008)}]{Petreczky:2007js}
\bibinfo{author}{\bibfnamefont{P.}~\bibnamefont{Petreczky}},
  \bibinfo{journal}{J. Phys.} \textbf{\bibinfo{volume}{G35}},
  \bibinfo{pages}{044033} (\bibinfo{year}{2008}), \eprint{0710.5561}.

\bibitem[{\citenamefont{Meyer}(2007)}]{Meyer:2007ic}
\bibinfo{author}{\bibfnamefont{H.~B.} \bibnamefont{Meyer}},
  \bibinfo{journal}{Phys. Rev.} \textbf{\bibinfo{volume}{D76}},
  \bibinfo{pages}{101701} (\bibinfo{year}{2007}), \eprint{0704.1801}.

\bibitem[{\citenamefont{Aharony et~al.}(2000)\citenamefont{Aharony, Gubser,
  Maldacena, Ooguri, and Oz}}]{Aharony:1999ti}
\bibinfo{author}{\bibfnamefont{O.}~\bibnamefont{Aharony}},
  \bibinfo{author}{\bibfnamefont{S.~S.} \bibnamefont{Gubser}},
  \bibinfo{author}{\bibfnamefont{J.~M.} \bibnamefont{Maldacena}},
  \bibinfo{author}{\bibfnamefont{H.}~\bibnamefont{Ooguri}}, \bibnamefont{and}
  \bibinfo{author}{\bibfnamefont{Y.}~\bibnamefont{Oz}}, \bibinfo{journal}{Phys.
  Rept.} \textbf{\bibinfo{volume}{323}}, \bibinfo{pages}{183}
  (\bibinfo{year}{2000}), \eprint{hep-th/9905111}.

\bibitem[{\citenamefont{Kovtun et~al.}(2005)\citenamefont{Kovtun, Son, and
  Starinets}}]{Kovtun:2004de}
\bibinfo{author}{\bibfnamefont{P.}~\bibnamefont{Kovtun}},
  \bibinfo{author}{\bibfnamefont{D.~T.} \bibnamefont{Son}}, \bibnamefont{and}
  \bibinfo{author}{\bibfnamefont{A.~O.} \bibnamefont{Starinets}},
  \bibinfo{journal}{Phys. Rev. Lett.} \textbf{\bibinfo{volume}{94}},
  \bibinfo{pages}{111601} (\bibinfo{year}{2005}), \eprint{hep-th/0405231}.

\bibitem[{\citenamefont{Myers et~al.}(2009)\citenamefont{Myers, Paulos, and
  Sinha}}]{Myers:2008yi}
\bibinfo{author}{\bibfnamefont{R.~C.} \bibnamefont{Myers}},
  \bibinfo{author}{\bibfnamefont{M.~F.} \bibnamefont{Paulos}},
  \bibnamefont{and} \bibinfo{author}{\bibfnamefont{A.}~\bibnamefont{Sinha}},
  \bibinfo{journal}{Phys. Rev.} \textbf{\bibinfo{volume}{D79}},
  \bibinfo{pages}{041901} (\bibinfo{year}{2009}), \eprint{0806.2156}.

\bibitem[{\citenamefont{Myers and Wapler}(2008)}]{Myers:2008me}
\bibinfo{author}{\bibfnamefont{R.~C.} \bibnamefont{Myers}} \bibnamefont{and}
  \bibinfo{author}{\bibfnamefont{M.~C.} \bibnamefont{Wapler}},
  \bibinfo{journal}{JHEP} \textbf{\bibinfo{volume}{12}}, \bibinfo{pages}{115}
  (\bibinfo{year}{2008}), \eprint{0811.0480}.

\bibitem[{\citenamefont{Danielewicz and Gyulassy}(1985)}]{Danielewicz:1984ww}
\bibinfo{author}{\bibfnamefont{P.}~\bibnamefont{Danielewicz}} \bibnamefont{and}
  \bibinfo{author}{\bibfnamefont{M.}~\bibnamefont{Gyulassy}},
  \bibinfo{journal}{Phys. Rev.} \textbf{\bibinfo{volume}{D31}},
  \bibinfo{pages}{53} (\bibinfo{year}{1985}).

\bibitem[{\citenamefont{Kats and Petrov}(2009)}]{Kats:2007mq}
\bibinfo{author}{\bibfnamefont{Y.}~\bibnamefont{Kats}} \bibnamefont{and}
  \bibinfo{author}{\bibfnamefont{P.}~\bibnamefont{Petrov}},
  \bibinfo{journal}{JHEP} \textbf{\bibinfo{volume}{01}}, \bibinfo{pages}{044}
  (\bibinfo{year}{2009}), \eprint{0712.0743}.

\bibitem[{\citenamefont{Buchel et~al.}(2009)\citenamefont{Buchel, Myers, and
  Sinha}}]{Buchel:2008vz}
\bibinfo{author}{\bibfnamefont{A.}~\bibnamefont{Buchel}},
  \bibinfo{author}{\bibfnamefont{R.~C.} \bibnamefont{Myers}}, \bibnamefont{and}
  \bibinfo{author}{\bibfnamefont{A.}~\bibnamefont{Sinha}},
  \bibinfo{journal}{JHEP} \textbf{\bibinfo{volume}{03}}, \bibinfo{pages}{084}
  (\bibinfo{year}{2009}), \eprint{0812.2521}.

\bibitem[{\citenamefont{Cai et~al.}(2009)\citenamefont{Cai, Nie, Ohta, and
  Sun}}]{Cai:2009zv}
\bibinfo{author}{\bibfnamefont{R.-G.} \bibnamefont{Cai}},
  \bibinfo{author}{\bibfnamefont{Z.-Y.} \bibnamefont{Nie}},
  \bibinfo{author}{\bibfnamefont{N.}~\bibnamefont{Ohta}}, \bibnamefont{and}
  \bibinfo{author}{\bibfnamefont{Y.-W.} \bibnamefont{Sun}},
  \bibinfo{journal}{Phys. Rev.} \textbf{\bibinfo{volume}{D79}},
  \bibinfo{pages}{066004} (\bibinfo{year}{2009}), \eprint{0901.1421}.

\bibitem[{\citenamefont{Cherman et~al.}(2008)\citenamefont{Cherman, Cohen, and
  Hohler}}]{Cherman:2007fj}
\bibinfo{author}{\bibfnamefont{A.}~\bibnamefont{Cherman}},
  \bibinfo{author}{\bibfnamefont{T.~D.} \bibnamefont{Cohen}}, \bibnamefont{and}
  \bibinfo{author}{\bibfnamefont{P.~M.} \bibnamefont{Hohler}},
  \bibinfo{journal}{JHEP} \textbf{\bibinfo{volume}{02}}, \bibinfo{pages}{026}
  (\bibinfo{year}{2008}), \eprint{0708.4201}.

\bibitem[{\citenamefont{Cohen}(2007)}]{Cohen:2007qr}
\bibinfo{author}{\bibfnamefont{T.~D.} \bibnamefont{Cohen}},
  \bibinfo{journal}{Phys. Rev. Lett.} \textbf{\bibinfo{volume}{99}},
  \bibinfo{pages}{021602} (\bibinfo{year}{2007}), \eprint{hep-th/0702136}.

\bibitem[{\citenamefont{Son}(2008)}]{Son:2007xw}
\bibinfo{author}{\bibfnamefont{D.~T.} \bibnamefont{Son}},
  \bibinfo{journal}{Phys. Rev. Lett.} \textbf{\bibinfo{volume}{100}},
  \bibinfo{pages}{029101} (\bibinfo{year}{2008}), \eprint{0709.4651}.

\bibitem[{\citenamefont{Noronha-Hostler
  et~al.}(2009)\citenamefont{Noronha-Hostler, Noronha, and
  Greiner}}]{NoronhaHostler:2008ju}
\bibinfo{author}{\bibfnamefont{J.}~\bibnamefont{Noronha-Hostler}},
  \bibinfo{author}{\bibfnamefont{J.}~\bibnamefont{Noronha}}, \bibnamefont{and}
  \bibinfo{author}{\bibfnamefont{C.}~\bibnamefont{Greiner}},
  \bibinfo{journal}{Phys. Rev. Lett.} \textbf{\bibinfo{volume}{103}},
  \bibinfo{pages}{172302} (\bibinfo{year}{2009}), \eprint{0811.1571}.

\bibitem[{\citenamefont{Peshier and Cassing}(2005)}]{Peshier:2005pp}
\bibinfo{author}{\bibfnamefont{A.}~\bibnamefont{Peshier}} \bibnamefont{and}
  \bibinfo{author}{\bibfnamefont{W.}~\bibnamefont{Cassing}},
  \bibinfo{journal}{Phys. Rev. Lett.} \textbf{\bibinfo{volume}{94}},
  \bibinfo{pages}{172301} (\bibinfo{year}{2005}), \eprint{hep-ph/0502138}.

\bibitem[{\citenamefont{Langer}(1967)}]{Langer:1967ax}
\bibinfo{author}{\bibfnamefont{J.~S.} \bibnamefont{Langer}},
  \bibinfo{journal}{Ann. Phys.} \textbf{\bibinfo{volume}{41}},
  \bibinfo{pages}{108} (\bibinfo{year}{1967}).

\bibitem[{\citenamefont{Langer}(1969)}]{Langer:1969bc}
\bibinfo{author}{\bibfnamefont{J.~S.} \bibnamefont{Langer}},
  \bibinfo{journal}{Ann. Phys.} \textbf{\bibinfo{volume}{54}},
  \bibinfo{pages}{258} (\bibinfo{year}{1969}).

\bibitem[{\citenamefont{Arrizabalaga and
  Reinosa}(2007{\natexlab{a}})}]{Arrizabalaga:2006hj}
\bibinfo{author}{\bibfnamefont{A.}~\bibnamefont{Arrizabalaga}}
  \bibnamefont{and} \bibinfo{author}{\bibfnamefont{U.}~\bibnamefont{Reinosa}},
  \bibinfo{journal}{Nucl. Phys.} \textbf{\bibinfo{volume}{A785}},
  \bibinfo{pages}{234} (\bibinfo{year}{2007}{\natexlab{a}}),
  \eprint{hep-ph/0609053}.

\bibitem[{\citenamefont{Jakovac}(2007)}]{Jakovac:2006gi}
\bibinfo{author}{\bibfnamefont{A.}~\bibnamefont{Jakovac}},
  \bibinfo{journal}{Phys. Rev.} \textbf{\bibinfo{volume}{D76}},
  \bibinfo{pages}{125004} (\bibinfo{year}{2007}), \eprint{hep-ph/0612268}.

\bibitem[{\citenamefont{Arrizabalaga and
  Reinosa}(2007{\natexlab{b}})}]{Arrizabalaga:2007zz}
\bibinfo{author}{\bibfnamefont{A.}~\bibnamefont{Arrizabalaga}}
  \bibnamefont{and} \bibinfo{author}{\bibfnamefont{U.}~\bibnamefont{Reinosa}},
  \bibinfo{journal}{Eur. Phys. J.} \textbf{\bibinfo{volume}{A31}},
  \bibinfo{pages}{754} (\bibinfo{year}{2007}{\natexlab{b}}).

\bibitem[{\citenamefont{Peshkin and Schroeder}(1995)}]{PeshkinSchroeder}
\bibinfo{author}{\bibfnamefont{M.}~\bibnamefont{Peshkin}} \bibnamefont{and}
  \bibinfo{author}{\bibfnamefont{D.}~\bibnamefont{Schroeder}},
  \emph{\bibinfo{title}{An introductin to Quantum Field Theory}}
  (\bibinfo{publisher}{Westview Press}, \bibinfo{year}{1995}).

\bibitem[{\citenamefont{Bogoljubov and Shirkov}(1959)}]{BogShir}
\bibinfo{author}{\bibfnamefont{N.}~\bibnamefont{Bogoljubov}} \bibnamefont{and}
  \bibinfo{author}{\bibfnamefont{D.}~\bibnamefont{Shirkov}},
  \emph{\bibinfo{title}{Introduction to the theory of quantized fields}}
  (\bibinfo{publisher}{Interscience Publishers Ltd. London},
  \bibinfo{year}{1959}).

\bibitem[{\citenamefont{Biro et~al.}(2007)\citenamefont{Biro, Levai, Van, and
  Zimanyi}}]{Biro:2006iy}
\bibinfo{author}{\bibfnamefont{T.~S.} \bibnamefont{Biro}},
  \bibinfo{author}{\bibfnamefont{P.}~\bibnamefont{Levai}},
  \bibinfo{author}{\bibfnamefont{P.}~\bibnamefont{Van}}, \bibnamefont{and}
  \bibinfo{author}{\bibfnamefont{J.}~\bibnamefont{Zimanyi}},
  \bibinfo{journal}{Phys. Rev.} \textbf{\bibinfo{volume}{C75}},
  \bibinfo{pages}{034910} (\bibinfo{year}{2007}), \eprint{hep-ph/0606076}.

\bibitem[{\citenamefont{Blaizot and Iancu}(1997)}]{Blaizot:1996az}
\bibinfo{author}{\bibfnamefont{J.-P.} \bibnamefont{Blaizot}} \bibnamefont{and}
  \bibinfo{author}{\bibfnamefont{E.}~\bibnamefont{Iancu}},
  \bibinfo{journal}{Phys. Rev.} \textbf{\bibinfo{volume}{D55}},
  \bibinfo{pages}{973} (\bibinfo{year}{1997}), \eprint{hep-ph/9607303}.

\end{thebibliography}

\end{document}